\documentclass[prc,aps,tightenlines,floatfix,showpacs,twocolumn]{revtex4}
\usepackage[dvips]{graphics,color}
\usepackage{graphicx}
\usepackage{dcolumn}
\usepackage{bm}
\usepackage{epsfig}
\usepackage{supertabular}
\usepackage{amsmath}

\newcommand{\beq}{\begin{equation}}
\newcommand{\eeq}{\end{equation}}
\newcommand{\beqa}{\begin{eqnarray}}
\newcommand{\eeqa}{\end{eqnarray}}
\newcommand{\bd}[1]{ \mbox{\boldmath $#1$}}

\begin{document}
\def\ii{\'\i}

\title{Phenomenological and microscopic cluster models II. Phase transitions
}

\author{ P. R. Fraser$^1$,
H. Y\'epez-Mart\ii nez$^2$,
P. O. Hess$^1$
and G. L\'evai$^3$  \\
{\small\it
$^1$ Instituto de Ciencias Nucleares, UNAM, Circuito Exterior, C.U.,} \\
{\small\it A.P. 70-543, 04510 M\'exico, D.F., Mexico} \\
{\small\it
$^2$Universidad Aut\'onoma de la Ciudad de M\'exico,
Prolongaci\'on San Isidro 151,} \\
{\small\it
Col. San Lorenzo Tezonco, Del. Iztapalapa,
09790 M\'exico D.F., Mexico} \\
{\small\it
$^3$Institute of Nuclear Research of the
Hungarian Academy of Sciences,} \\
{\small\it
Debrecen, Pf. 51, Hungary-4001} \\
}

\begin{abstract}
 Based on the results of a previous paper (Paper I), by performing the
 geometrical mapping via coherent states, phase transitions are
 investigated and compared within two algebraic cluster models.  The
 difference between the Semimicroscopic Algebraic Cluster Model (SACM)
 and the Phenomenological Algebraic Cluster Model (PACM) is that the
 former strictly observes the Pauli exclusion principle between
 the nucleons of the individual clusters, while the latter ignores
 it. From the technical point of view the SACM is more involved
 mathematically, while the formalism of the PACM is closer to that of
 other algebraic models with different physical content. First- and
 second-order phase transitions are identified in both models, while
 in the SACM a critical line also appears. Analytical results are
 complemented with numerical studies on $\alpha$-cluster states of the
 $^{20}$Ne and $^{24}$Mg nuclei.
\end{abstract}

\maketitle

\section{Introduction}

In a former contribution \cite{pap1}, called from now on Paper I, we
investigated the geometric mapping of the {\it Phenomenological
Algebraic Cluster Model} (PACM) and the {\it Semimicroscopic
Algebraic Cluster Model} (SACM) \cite{sacm1,sacm2}.  The first does
not observe the Pauli exclusion principle while the second one does.
The PACM belongs to the same family as models like the vibron model
{\cite{vibron}.  In both type of models we considered the same
Hamiltonian, while the model space of each is quite different. In
the SACM, observation of the Wildermuth condition \cite{wildermuth}
means that the number of relative oscillation quanta is restricted
from below while in the PACM the relative oscillation quanta start
from zero.

The method of geometrical mapping is gleaned from \cite{geom} for the
SACM, which reduces to the usual one
\cite{roosmalen1,roosmalen2,kirson} when no Pauli exclusion principle
is taken into account.  We showed in \cite{pap1} that the differences
in the mapped geometrical potential are large. However, within the PACM
one can reproduce the geometric potential of the SACM by including
very complicated higher-order interactions.

We also included in \cite{pap1} a discussion on how to define
correctly the coherent state parameter, such that for a large total
number of bosons ($N$, which plays the role of a cutoff) the mapped
potential will be independent of the cutoff $N$.

In this paper we concentrate on the study of phase transitions and
show that not only second-order phase transitions may occur, but also
first-order phase transitions. Furthermore, in the SACM a critical
line appears beyond which no phase transition occurs. Thus the
structure of the phase diagram of the SACM will be much richer.

We will also apply the mapping to two kind of systems, one with two
spherical clusters and the other one with a deformed and a spherical
cluster.  Numerical studies are added. We will show that the PACM
leads to inconsistencies when applied to real nuclei, while the SACM
performs well.

The paper is structured as follows: in section \ref{two} a general
discussion on phase transition, both in the PACM and SACM, is given,
independent of a particular system.  In Section \ref{three}, 
results are illustrated with two particular cluster systems, the
$\alpha$-cluster states of $^{20}$Ne and $^{24}$Mg.  In the former,
both clusters are spherical, while in the latter, one of them is
deformed, leading to a more complex physical situation.  Finally, in
Section \ref{four} conclusions are drawn and a discussion is presented
on the differences between the PACM and the SACM and their importance in the
study of nuclear clusters.

\section{Phase transitions in algebraic cluster models}
\label{two}

In what follows we apply the formalism developed in Paper I
\cite{pap1} to discuss phase transitions in the SACM and PACM. Before
that we present the general framework within which the discussion will
be implemented.

\subsection{Definition of a phase transition}
\label{two-one}

Phase transitions are investigated using the following steps and the
recommendations of \cite{greiner-thermo,octavio,octavio2}. This method
can be applied to any system and does not depend on the notion of
symmetries.  This presentation does not need the language of
catastrophe theory \cite{gilmore}, which simplifies considerations.

\noindent
a) In the first step the minima of the potential energy surface (PES) 
are determined in the space of the collective variables $\alpha_m$.  
In the present case there is only one relevant variable $\alpha$.
This is due to the fact that the distance vector between the clusters
can always be aligned along the intrinsic $z$-axis which connects the two 
clusters. The extrema are obtained from
\begin{eqnarray}
\frac{{\rm d} V}{{\rm d}\alpha} & = 0 ~~~.
\end{eqnarray}
This determines the position of the extrema at\linebreak ${\bar \alpha}_i$
($i=1,2,..$) and the values $V({\bar \alpha}_i)$ of the potential
there.  The ${\bar \alpha}_i$ are the values of the variable $\alpha$
at the $i^{\text{th}}$ minimum, which is a function of the interaction
parameters ${\vec p}=(p_k)$, with $p_k$ as a short-hand notation for
the parameters $k=1,2,...,n_k$, with $n_k$ being the number of
parameters.
\\ b) Once the extrema are obtained, one determines {\it at each
minimum} the first and second derivatives of the potential with
respect to the parameters of the model, i.e.,
\begin{eqnarray}
\frac{\partial^n V({\bar \alpha}_i)}{\partial p_k^n}\ , ~~n=1,2~~~.
\end{eqnarray}
When we discuss the PACM and SACM further below, we will give first a
general discussion on the properties of phase transitions, independent
of the values of a particular subset of interaction parameters.  In
the subsequent concrete applications, however, we will fix most
interaction parameters and vary only $x$ and $y$, 
which control the transition from one
effective symmetry \cite{huitz1} to another.
This will give us
particular curves in the space of phase transitions.  
\\
c) The locations of phase transitions are determined by identifying the 
borders in the parameter space where at least two minima are at
equal energy, i.e.,
\begin{eqnarray}
V({\bar \alpha}_{i_1}) & = & V({\bar \alpha}_{i_2})
\end{eqnarray}
for some index values $i_1$ and $i_2$ of the minima.
This results in a relation
\begin{eqnarray}
f(p_1, ..., p_{n_p}) & = & 0
\end{eqnarray}
between the parameters, which allows one, in principle, to express one
parameter in terms of the others (see the particular examples
further below).
\\
d) The phase transition is of order $m$ when, up to $n=m-1$, the
derivatives of the potential with respect to the free parameters are
equal at the point of phase transition, while the $m^{\text{th}}$ derivative of
the potential, with respect to its parameters, is discontinuous at the
point of the phase transition, i.e.,
\begin{eqnarray}
\frac{\partial^n V({\bar \alpha}_{i_1})}{\partial p_k^n} & = &
\frac{\partial^n V({\bar \alpha}_{i_2})}{\partial p_k^n}
~~~,~~~ {\rm for}~ n<m
\nonumber \\
\frac{\partial^m V({\bar \alpha}_{i_1})}{\partial p_k^m} & \ne &
\frac{\partial^m V({\bar \alpha}_{i_2})}{\partial p_k^m} 
~~~.
\end{eqnarray}

Note again that this procedure for determining the order of a phase
transition is quite general and does not depend on identifying the
phase with a dynamical symmetry. Thus, this procedure can also be applied
to systems that do not exhibit dynamical symmetries.

\subsection{A possible explanation for the differences in using coherent states
and a numerical approach}
\label{three-two}

In the numerical calculations further below, we will see that, though the
use of coherent states shows a clear phase transition, this is not the
case in the numerical study, because the total number of bosons
$N$ is fixed. In what follows, we indicate why this is so.

The main reason for using coherent states is that the phase transition
is given by the crossing of two potential minima. In the case of a
first-order phase transition, the two minima are well separated by a
barrier and a simple crossing takes place. For the case of a 
second-order phase transition, the spherical minimum coincides with the
``deformed" one at $\alpha = 0$ at a certain point in the parameter
space. Once one minimum is deeper than the other one, an immediate
jump by hand to the global minimum is made, using coherent states.
This appears to produce the well defined phase transition. However, in
numerical calculations within a finite system, the states are a mixture
of states in both minima. Around the point of phase transition, the
eigenstates of the Hamiltonian are a mixture of states in {\it both}
minima. Only far away the states are located clearly in one or the
other minimum.  This produces, for finite $N$, a gradual change in the
wave function and as a consequence a gradual change of the control
parameters, like the energy or the expectation value of
$\boldsymbol{n}_\pi$, which define the phase transition. For\linebreak
$N\rightarrow\infty$, the discontinuities in the derivatives of the
potential will be clear cut.

As a consequence, the apparently clear phase transition for any number
of $N$ is not a real one, because $N$ is finite.
Nevertheless, the use of coherent states shows
the points of phase transitions and of which order they will be for
$N\rightarrow\infty$.

\subsection{Study of phase transitions in the SACM}

Paper I \cite{pap1} contains the general expression of the Hamiltonian 
(in subsection 2.2) and the geometrically mapped SACM potential 
(in subsection 4.1). Here we recall only the essential formula, 
Eq.~(30) of Paper I, necessary for the present discussion: 
\begin{eqnarray}
\langle \mbox{\boldmath$H$}\rangle &=&
{\cal C}(x,y)-(b+{\bar b})xy 
\left(
A(x,y)\alpha^2\frac{F_{11}\left( \alpha ^{2}\right) }{%
F_{00}\left( \alpha ^{2}\right) }
\right.
\nonumber \\
&&\left.
-B(x,y)\alpha^4\frac{%
F_{22}\left( \alpha ^{2}\right) }{F_{00}\left( \alpha ^{2}\right) } 
+\alpha^6\frac{F_{33}\left( \alpha ^{2}\right) }{%
F_{00}\left( \alpha ^{2}\right) }
\right.
\nonumber \\
&&\left.
-C(x,y)\alpha^2\frac{%
F_{20}^{N-2}\left( \alpha ^{2}\right) }{F_{00}\left( \alpha ^{2}\right) }%
\right)
\label{exp-h}
\end{eqnarray}
We also note that the $x=0$ case relevant to the $SO(4)$ to $SO(3)$
phase transition has to be discussed separately. (See Eqs.~(35) to
(38) in Paper I \cite{pap1}.)

\subsubsection{Phase transition diagram for the SACM:
General considerations}
\label{two-two-one}

The complex structure of the geometrically mapped potential
complicates an analytic treatment of the problem. It is therefore
essential to formulate a set of criteria facilitating a
straightforward way to determine the order of phase transitions.

Equation (\ref{exp-h}) demonstrates that the explicit dependence of the
potential on the parameters $A$, $B$ and $C$ is linear of the type
\begin{eqnarray}
{\widetilde V} & = & \sum_k p_k \alpha^{m_k}f_k(\alpha )
~~~,
\label{v-tilde}
\end{eqnarray}
with $m_k>1$ and $p_k$ being a short-hand notation for the $k^{\text{th}}$
parameter. The $f_k$ are given by ratios of the functions
$F_{pq}(\alpha )$ and are always greater than zero for $\alpha\ne
0$. Only the function $F_{20}^{N-2}$ approaches zero for
$\alpha\rightarrow\infty$. (See Eq.~(39) in Paper I \cite{pap1}.)

According to the general discussion on phase transitions in Subsection
\ref{two-one}, let us now turn to the potential minima located at
${\bar \alpha}_i$. We investigate their dependence on the parameters
$p_k$, standing for $A$, $B$ and $C$. The structure of
(\ref{v-tilde}) guarantees that ${\bar \alpha}_1=0$ is always an
extremum. Furthermore, not only its first-order derivatives, but the
function value is also zero at ${\bar \alpha}_1=0$.  It is thus
sufficient to focus on the second, deformed minimum, with ${\bar
\alpha}_2$, for which the following consideration holds. There are
two possibilities: \\
a) ${\bar \alpha}_2>0$, or \\
b) ${\bar \alpha}_2=0$. \\
The consequences are seen by determining the first absolute derivative
of the potential with respect to the parameter $p_k$.  This first
derivative is given by
\begin{eqnarray}
\frac{{\rm d}{\widetilde V}}{{\rm d}p_k} & = &
\frac{\partial{\widetilde V}}{\partial p_k}
+ \frac{\partial {\widetilde V}}{\partial{\bar \alpha}_i}
\frac{\partial{\bar \alpha}_i}{\partial p_k}
\nonumber \\
& = & \frac{\partial{\widetilde V}}{\partial p_k}
~~~,
\end{eqnarray}
because $\frac{\partial {\widetilde V}}{\partial{\bar \alpha}_i}$
vanishes at the minimum.  Taking into account (\ref{v-tilde}), this is
further expressed as
\begin{eqnarray}
\frac{{\rm d}{\widetilde V}}{{\rm d}p_k} 
& = & {\bar \alpha}_2^{m_k}f_k({\bar \alpha}_2)
~~~.
\label{first}
\end{eqnarray}

For case a) this expression is different from zero (remember that
${\bar \alpha}_2>0$), while the derivative within the spherical
minimum is equal to zero. In other words, there is a {\it first-order}
phase transition.

For case b), (\ref{first}) is equal to zero, and therefore the phase
transition must be of higher order. To determine which order,
second-order derivatives are also needed:
\begin{eqnarray}
\frac{{\rm d}^2 {\widetilde V}}{{\rm d}p_k^2} 
& = & \frac{\rm d}{{\rm d}p_k}\left(
\frac{\partial {\widetilde V}}{\partial p_k} \right)
\nonumber \\
& = & \frac{\partial {\bar \alpha}_2^{m_k}}{\partial p_k}f_k({\bar \alpha}_2)
+ {\bar \alpha}_2^{m_k}\frac{\partial f_k({\bar \alpha}_2)}{{\bar \alpha}_2}
\frac{\partial {\bar \alpha}_2}{\partial p_k}
~~~.
\label{vpp}
\end{eqnarray}
Here we used the fact that in (\ref{first}) there is no explicit
dependence in $p_k$ left. Since in this case ${\bar \alpha}_2=0$, 
(\ref{vpp}) reduces to 
\begin{eqnarray}
\frac{{\rm d}^2 {\widetilde V}}{{\rm d}p_k^2} & = &
\frac{\partial {\bar \alpha}_2^{m_k}}{\partial p_k}f_k({\bar \alpha}_2)
~~~.
\label{d2v}
\end{eqnarray}
As both the $f_k$ functions and the partial derivatives are different
from zero, so is (\ref{d2v}) in general, i.e.  the phase transition is
of {\it second order} in case b).

In conclusion we have the simple identification of phase transitions:\\
a) When the deformed solution ${\bar \alpha}_2$ is {\it different from zero},
then there is a {\it first-order} phase transition. \\
b) When the deformed solution ${\bar \alpha}_2$ is {\it equal zero}, 
then there is a {\it second-order} phase transition.

\subsubsection{General discussion of phase transitions in the SACM}

In Fig.~\ref{fig01} the surface where a phase transition takes place
is plotted in the space of the independent parameters $A$, $B$ and
$C$.  The solid line marks a change from first-order to second-order
transitions.  For larger, positive $C$ the transition is of second
order, while for smaller, negative $C$ it is of first order. For
completeness we show in Fig.~\ref{fig02} the phase space diagram for
the case when $x=0$, as it is the case for the $SO(4)$ to $SO(3)$
transition.

A remarkable finding is that at approximately\linebreak $C \approx
-15$ the surface of phase transition ceases to exist. Beyond that
point, no phase transition can be observed, i.e., the straight
dashed line in Fig.~\ref{fig01} at
approximately $C \approx -15$ represents a {\it critical line}.  This
is shown by a dashed line in Fig.~\ref{fig01}. One then can trace a
straight line from below the surface, around the critical line, ending
up above the surface without passing through a phase transition, which
is similar to the critical point in the two-dimensional phase diagram
of water.  Fixing all interaction parameters except $x$ and $y$ (as we
will do in the numerical applications), passing from one dynamical
symmetry limit to another one will trace a line in this
three-dimensional space. Depending on the fixed parameters, this line
will or will not cross the surface of phase transition.

\begin{figure}[htp]
\begin{center}
\scalebox{0.5}{\includegraphics*{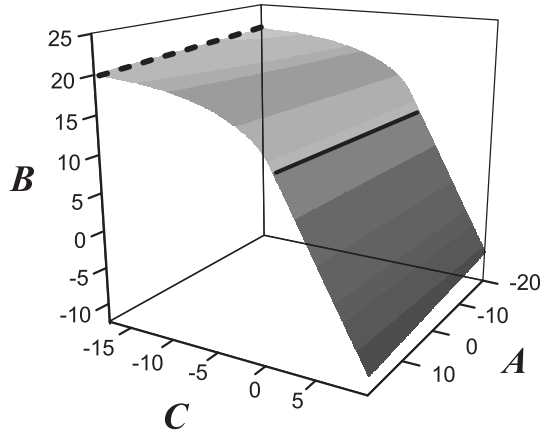}}
\end{center}
\caption{ \label{fig01}
The phase space diagram of the SACM as a function of the independent
parameters $A$, $B$ and $C$. The solid line marks the change from a
second- to a first-order phase transition, while the dashed
line represents a `critical line'.}
\end{figure}

\begin{figure}[htp]
\centerline{\epsfxsize=9.5cm\epsffile{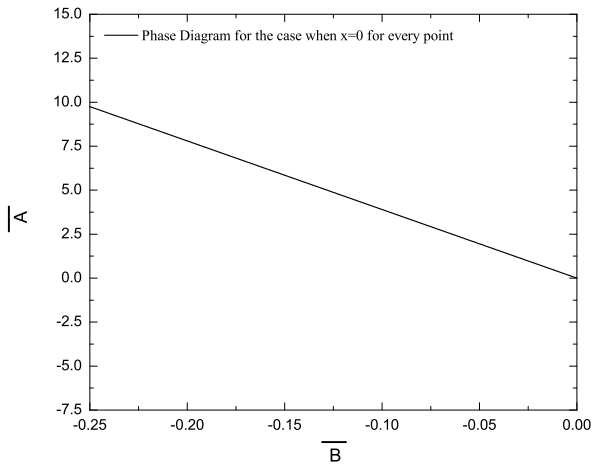}}
\caption{ \label{fig02}
The phase space diagram as a function of the independent parameters
$\overline{A}$ and $\overline{B}$ for the $SO(4)$ to $SO(3)$
transition, within the SACM.
As defined in Paper I, within the $SO(4)$ dynamical limit the $x=0$
and one has to define a new representation of the potential, involving
the new parameters $\overline{A}$ and $\overline{B}$. 
(See also Paper I \cite{pap1} for
more.)  }
\end{figure}

\begin{figure}[htp]
\centerline{\epsfxsize=6.5cm\epsffile{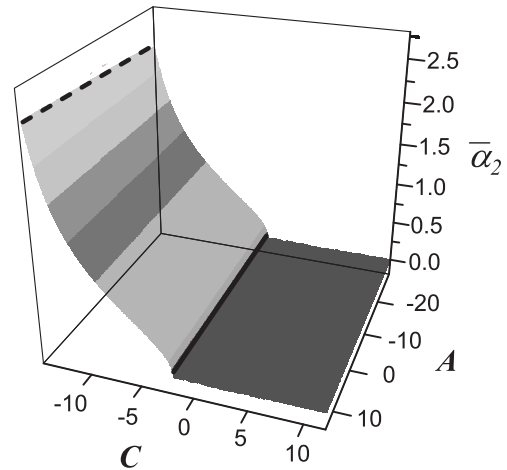}}
\caption{ \label{fig03} 
The variable ${\bar \alpha}_2$ of the deformed solution, as a function
in $A$ and $C$. $B$ is fixed by the requirement that one is at a point
of a phase transition. The solid line marks the transition from a
second- to a first-order phase transition, the dashed a `critical line'.}
\end{figure}

Figure \ref{fig03} displays ${\bar \alpha}_2$ corresponding to the
deformed solution {\it at the point of phase transition} as a function
in $A$ and $C$. 
(Remember that the {\it spherical solution}
corresponds to the always existing extremum at $\alpha = 0$ in the
case it is a local minimum.) 
Each point of the surface represents
also a given $B$ at which the phase transition occurs, i.e., $B$ is
fixed by the requirement that there is a phase transition.  In this
figure the solid line also represents the change from one type of
phase transition to the other one. For larger, positive $C$ ${\bar
  \alpha}_2 = 0$ holds, and according to the discussion of subsection
\ref{two-two-one}, it corresponds to a second-order phase transition.
For smaller, negative $C$ the ${\bar \alpha}_2>0$ holds at the point
of phase transition, i.e., it corresponds to a first-order phase
transition.

The whole phase
structure of the system is illustrated in Fig.~\ref{fig6}, which
corresponds to the fixed value $A=10$.  The solid line represents a
cut through the phase transition surface.  For $C>0$ the phase
transition is of second order, while for $C<0$ is of first order. The
Roman numbers indicate the following regions: i) Region I corresponds
to the existence of two minima (one spherical and the other deformed),
with the deformed one as the global minimum. ii) Region II also
corresponds to two minima with the spherical as the global one.  iii)
In region III there is only one spherical minimum while iv) in region
IV there is only one deformed minimum.  The region denoted by a zero
refers to potential with no minimum.  The other dashed lines do not
indicate phase transitions, rather they separate the areas where two
minima exist and the ones where only one minimum exists.  The line of
phase transition ends at approximately $C \approx -15$, corresponding
in this case to a point.

There is another point on the solid line, where the
first-order phase transition turns over into a second-order phase
transition.  A distinction has to be made between this and the
critical point of the phase diagram {\it of water}, where before the
critical point the phase transition is of first order, while {\it at}
the critical point it is of second order.  In our case, however, the
phase transition does not end at this point, rather it
continues as second-order one in the domain with $C>0$. One can now
draw a curve which encircles this line, going from a spherical phase
to a deformed one.  The value of $C$ at which this phenomenon happens,
varies with $A$ and $B$.

\begin{figure}[htp]
\begin{center}
\scalebox{0.33}{\includegraphics*{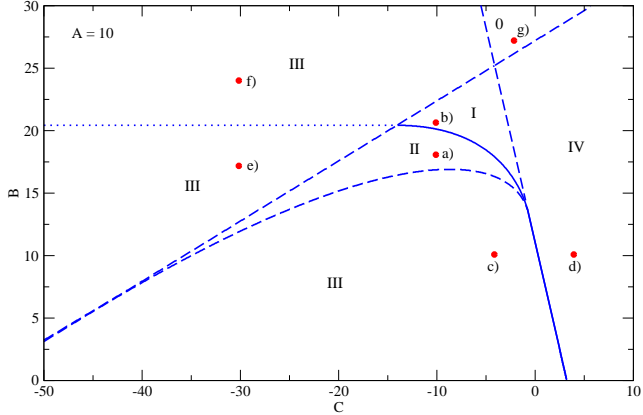}}
\end{center}
\caption{ \label{fig04}
(color online).
A cut through the phase at $A=10$. The notation is explained in the text.} 
\label{fig6}
\end{figure}

\begin{figure}[htp]
\begin{center}
\scalebox{0.38}{\includegraphics*{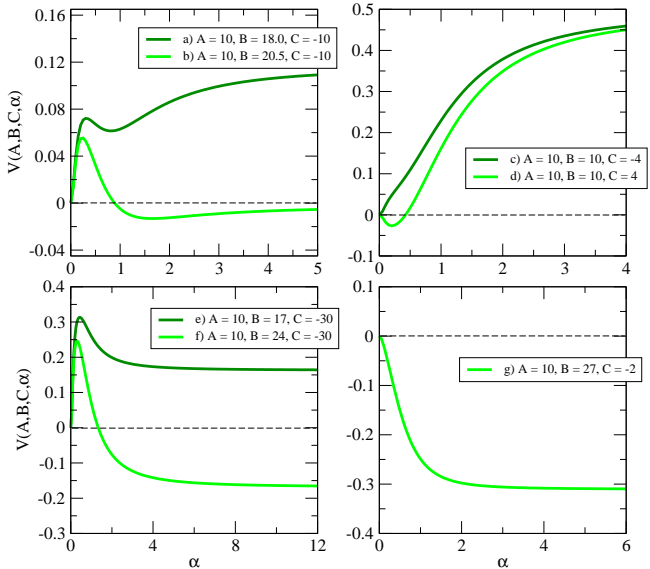}}
\end{center}
\caption{ \label{fig05}
(color online).
Several potentials for different values of $B$ and $C$, for $A=10$
are shown. Some potentials approach a positive value, i.e., for $N
\rightarrow \infty$ they goes to $+\infty$. Others approach a negative
value, i.e., for $N \rightarrow \infty$ they approach $-\infty$.  }
\end{figure}

\begin{figure}[htp]
\begin{center}
\scalebox{0.33}{\includegraphics*{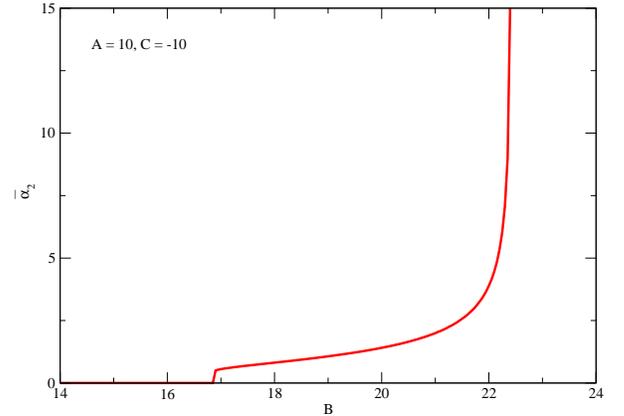}}
\end{center}
\caption{ \label{fig06}
(color online).
The value of ${\bar \alpha}_2$ at $A = 10$ and $C = -10$ as $B$ moves through regions.}
\end{figure}

The horizontal dotted line represents a division (see discussion in
\ref{two-one}): above this line the potential approaches a negative
infinite value for $N\rightarrow\infty$, i.e., the potential gets
unstable and the cluster system dissolves, once getting to the
``deformed'' solution. Below that line the potential approaches plus
infinity for $N\rightarrow\infty$, thus the deformed solution
corresponds to a stable cluster system.  

Fig.~\ref{fig05} gives a sample of energy functionals at a fixed $N$
($n_0 = 8$ and $N+n_0 = 20$) in different regions, the specific points
being marked with lower-case letters in Fig.~\ref{fig04}. The upper
left panel shows functionals either side of the first-order phase
transitions, points a) and b).  The upper right one shows graphs either
side of the second-order phase transitions, points c) and d). The
lower left panel shows a functional either side of the dotted line in the
Region III of greater $B$, points e) and f), showing that while each
have a barrier, one is bound as $\alpha \rightarrow \infty$, and the
other is unbound. These tails are dependent on $N$, with the unbound
approaching infinity and the bound negative infinity as $N \rightarrow
\infty$. Thus, this region is unphysical. The Region III of lesser
$B$, shown in the second panel, has no barrier, and is bound as
$\alpha \rightarrow \infty$. The fourth panel shows a fully unbound
potential with no minima from Region 0, point g).

For completeness, Fig.~\ref{fig06} traces the value of ${\bar \alpha}_2$
for\linebreak $14 < B < 24$ at $C = -10$ in Fig.~\ref{fig04}, which
thus represents a movement between several regions in that figure (and
is therefore analogous to being a subset of Fig.~\ref{fig03}). It shows that the critical
point corresponds to ${\bar \alpha}_2$ asymptotically approaching $\infty$ at
$B$ slightly greater than 22.

Note that the discussion of phase transitions is completely
independent of the system considered, which is of great advantage.
In the next section we will consider particular systems and study the
properties of their phase transitions.

\subsection{The PACM case: Pauli principle not taken into account}
\label{four-two}

Here we apply the results obtained in subsection 4.2 of Paper I \cite{pap1}. 
We display only the expression of the normalized potential in Eq.~(46) 
of Paper I: 
\begin{eqnarray}
{\widetilde V} & = & \left\{ A\beta^2 -B \beta^4 + \beta^6 \right\}
~~~.
\label{vtilde2}
\end{eqnarray}
We also remind the reader that the $SO(4)$ to $SO(3)$ transition needs
a separate discussion, because in that case the potential reduces to a
quartic one. (See the discussion in subsection 4.2 of Paper I
\cite{pap1}.)

\subsubsection{Phase transitions}

The extrema are found by setting the first derivative with respect to
$\beta$ equal to zero. We obtain, in general, two solutions
\begin{eqnarray}
{\rm solution~ 1:}~~~\beta_1 & = & 0
~~~,
\end{eqnarray}
noting that in this case the potential value is always
${\widetilde V}(\beta_1 )=0$, and
\begin{eqnarray}
{\rm solution~ 2:}~~~\beta_2^2 & = & \frac{1}{3} \left(
B \pm \sqrt{ B^2 - 3A } \right)
~~~,
\label{min2}
\end{eqnarray}
where the value of the potential at the extremum with the positive sign
is given by
\begin{multline}
{\widetilde V}(\beta_2 ) = \frac{1}{27}\left( B +
\sqrt{B^2 - 3A} \right) \\ \times \left( 6A-B^2-B\sqrt{B^2-3A}\right) 
~~~.
\end{multline}
The barrier maximum is obtained by choosing the negative sign in front
of the square root expressions.  The last equation is acceptable only
when the solution in (\ref{min2}) is real and positive.  The solution
is real for $A \le B^2/3$.  When $B<0$ and $A>0$ holds, no deformed
solution exist. In fact, inspecting Eq. (\ref{vtilde2}) for $A>0$
there is always a minimum at $\beta = 0$, while a negative $B$ always
gives a positive contribution, $+\beta^4$. When $A$ and $B$ are
negative, the potential is negative at small $\beta$ and is turned
over by the positive contributions $\sim +\beta^4$ and $+\beta^6$. In
this case, only a deformed minimum exists.

\begin{figure}[htp]
\begin{center}
\scalebox{0.35}{\includegraphics*{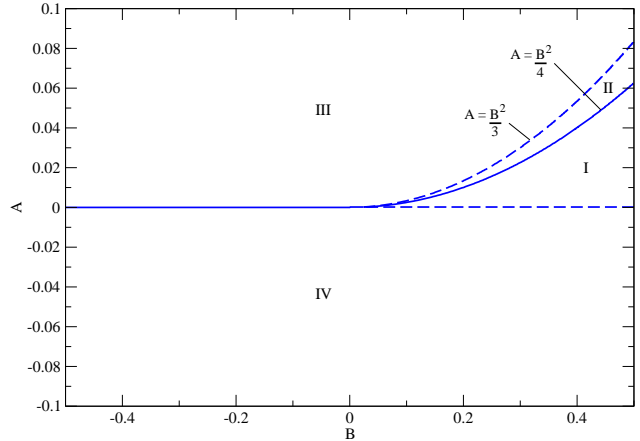}}
\end{center}
\caption{ \label{fig07}
(color online).
The parameter phase diagram for the
PACM~\cite{lorena-T,cocoyoc-2011}. The horizontal axis is $B$, while
the vertical axis corresponds to $A$. In Region I and 
Region II two minima exist, one spherical and one deformed. In 
Region I the global minimum is the deformed one, while in 
Region II it is the spherical minimum. In Region III only a
spherical minimum exists and in Region IV the only minimum is a
deformed one. }
\end{figure}

The structure explained above is summarized in Fig.~\ref{fig07}. The
horizontal line is the $B$ axis while the vertical one is the $A$
axis.  On the left hand side ($B<0$) a spherical minimum exists for
$A>0$, while for $A<0$ the minimum is deformed.  The phase transition
takes place at the line $A=0$. This will be discussed further below as
case b). For $B>0$ and $A<0$, a deformed minimum always exists. For
$A>0$ and below the dashed curve $A=B^2/3$ a deformed minimum
coexists with a spherical minimum until $A=0$ is reached.  The solid
curve is determined by requiring that the potential minima at
$\beta_1$ and $\beta_2$ are degenerate, i.e.,
\begin{eqnarray}
{\widetilde V}(\beta_1) & = & {\widetilde V}(\beta_2) ~~~,
\end{eqnarray}
which leads to the condition
\begin{eqnarray}
A & = & \frac{B^2}{4} ~~~.
\end{eqnarray}
Between the upper dashed and the solid curves the spherical minimum is
the global one, while below the solid curve the global minimum is the
deformed one. Crossing the solid line a phase transition takes place,
which will be called below case a). It is of no surprise that the
phase diagram in Fig.~\ref{fig07} is similar to Fig. 2 of
\cite{cejnar}.

\vskip 0.5cm
\noindent
{\bf Case a), $B>0$}:

The potential is exactly zero at $\beta_1=0$, and thus
\begin{eqnarray}
\frac{\partial^n {\widetilde V}(\beta_1=0)}{\partial A^n} & = & 0
\end{eqnarray}
for all $n \ge 0$.  Contrary to this, the first derivative of the
potential at $\beta_2$, the deformed minimum, taken at $A=B^2/4$
where the height of the deformed minimum equals that of the spherical
one, is given by
\begin{eqnarray}
\frac{\partial {\widetilde V}(\beta_2)}{\partial A} & = & \frac{1}{2}B ~~~,
\end{eqnarray}
which is clearly different from zero for $B>0$.  Thus the phase
transition is of {\it first order} for $B>0$.

We can repeat the calculation for the derivatives in $B$, leading
to the same conclusion. Also, any combination of curves given by
the variable $\left(c A - s B\right)$
(with $c={\rm cos}\phi$ and $s={\rm sin}\phi$, with $\phi$ as the angle
between the tilted straight line and the $A$ axis) 
leads to the same conclusion.

\vskip 0.5cm
\noindent
{\bf Case b): $B \le 0$}

We will vary $A$ from positive to negative values for a fixed $B$.
The same steps as taken for case a) can be applied, setting $B=-|B|$.
The first and second derivatives of the potential with respect to $A$
are given by
\begin{eqnarray}
\frac{\partial {\widetilde V}(\beta _2)}{\partial A}&=&
\frac{1}{3}\sqrt{B^2 - 3A} \left( 1- \frac{\mid B \mid}{\sqrt{B^2-3A}}\right)
\nonumber  \\
\frac{\partial^2 {\widetilde V}(\beta _2)}{\partial A^2} &=&
-\frac{1}{2\sqrt{B^2 - 3A}}
~~~.
\nonumber \\
\end{eqnarray}
Evaluating these derivatives at $A=0$ and $B=-|B|$ gives
\begin{eqnarray}
\frac{\partial {\widetilde V}}{\partial A} & = & 0 ~~~,~~~
\frac{\partial^2 {\widetilde V}}{\partial A^2} ~ = ~ -\frac{1}{2B}
~~~.
\end{eqnarray}
For this case we also have to determine the derivatives with respect
to $B$, given {\it at the point of phase transition}, i.e., setting
$A=0$ afterward, and $B=-\mid B\mid$,
\begin{eqnarray}
\frac{\partial {\widetilde V}}{\partial B} & = & -\frac{1}{9}
\left[ -|B| + \sqrt{|B|^2-3A}\right]^2~=~0 \nonumber \\
\frac{\partial^2 {\widetilde V}}{\partial B^2} & = & -\frac{2}{9}
\left[ -|B| + \sqrt{|B|^2-3A}\right]\left[ 1 -
\frac{|B|}{\sqrt{|B|^2-3A}} \right] \nonumber \\
& = &0
~~~.
\end{eqnarray}
Changing $B$ corresponds only to moving {\it on the line of phase
transition}, which does not make sense, because one has to {\it
cross} the line of phase transition. However, a combination of
changing $A$ and $B$ corresponds to an inclined straight line,
crossing the line of phase transition.  This means that one has to
consider a general line given by the variable $C=cA-sB$, with
arbitrary $c={\rm cos}\phi$ and $s={\rm sin}\phi$, with $\phi$ being an
arbitrary rotation angle.  Determining at the point of phase
transition the first- and second-order derivatives of ${\widetilde V}$
with respect to $C$ leads to
\begin{eqnarray}
\frac{\partial {\widetilde V}}{\partial C} & \rightarrow & c
\frac{\partial {\widetilde V}}{\partial A}
\nonumber \\
\frac{\partial^2 {\widetilde V}}{\partial C^2} & \rightarrow & c^2
\frac{\partial^2 {\widetilde V}}{\partial A^2}
~~~.
\end{eqnarray}
Because the first derivative of the potential with respect to $A$ is
zero, while the second is different from zero {\it at the point of
phase transition}, we can claim also that the first derivative with
respect to $C$ is zero, while the second is different from zero. We
conclude that the phase transition is of second order.

\section{Numerical studies}
\label{three}

This section deals with two widely known cluster systems: $^{20}$Ne as
$^{16}$O+$\alpha$, where both clusters are spherical, and $^{24}$Mg as
$^{20}$Ne + $\alpha$, where one of them is deformed.

Since our aim is studying transitions from one particular dynamical
symmetry limit to another one, we first fix all the interaction
parameters except $x$ and $y$. The $\hbar\omega$ parameter was not
fixed, but rather it was chosen according to the harmonic oscillator
constant corresponding to the unified nucleus.  The fixed interaction
parameters are determined in such a way that in the dynamical symmetry
limits the spectrum appears with the same scale as the physical
measured one. The spectrum of the real nucleus would probably
correspond to a single point in the $(x,y)$ parameter space. However,
our aim is not reproducing the exact spectrum, rather to reach a
conceptual understanding of phase transitions when going from one
dynamical symmetry to another, and investigating the {\it conjecture
that a phase is defined by an effective symmetry}.  One alternative
method would be to adjust several cluster systems and to try to find a
series of systems which, for example, would cross the surface of phase
transitions at one point. We do not follow this way, rather postponing
it for later consideration.

\subsection{Two spherical clusters: $^{16}$O+$\alpha$ $\rightarrow$
$^{20}$Ne}
\label{three-one}

In this case the only degree of freedom is the radial motion, as the
clusters do not have an internal structure apart from the fact that
they are composed of fermions, which have to obey the Pauli exclusion
principle. Therefore the cluster representation is $(\lambda_C,\mu_C$)
= $(0,0)$. As described in Paper I \cite{pap1}, the $SO(3)$ limit does
not exist as an independent limit in this case (the $SO(3)$
Hamiltonian is a reduced version of the the $SU(3)$ Hamiltonian), so
the $y=1$ choice has to be made.  The only transition to consider is
thus between the $SU(3)$ and the $SO(4)$ limits.

Concerning the determination of the parameters, one has to take into
account that some parameters appear in both dynamical symmetry limits,
like $\gamma$ preceding $\boldsymbol{L}^2$ (see Eqs. (10) and (11) in
Paper I \cite{pap1}). We first determine the parameters in the $SU(3)$
dynamical symmetry limit, which fixes $\gamma$, and then we determine
the remaining parameter $c$, which appears in the $SO(4)$ dynamical
symmetry limit. The terms ${\cal C}_2(\lambda_C,\mu_C)$ and
$\boldsymbol{L}_C^2$ do not
contribute to the Hamiltonian in this case, so the corresponding
parameters are kept zero. Note that in this case
${\bd L}_R^2={\bd L}^2$.

\subsubsection{The SACM}
\label{three-one-one}

In the first step the parameters are adjusted within the $SU(3)$
limit, setting $x=1$ and $y=1$ and in the $SO(4)$ limit, setting $x=0$
and $y=1$. The parameters are depicted in Table \ref{Ne20-param}.  The
SACM yields reasonable results, because the ground-state band belongs
to $n_\pi=8$, $( \lambda , \mu )=(8,0)$, where $n_\pi=8$ corresponds
to the minimal number of relative oscillation quanta $n_0$ required by
the Wildermuth condition. The first excited $0^+$ state corresponds to
a 2$\hbar\omega$ excitation and naturally lies at high energy as
required by the experimental data. The spectra in the $SO(4)$ and
$SU(3)$ limits are shown in the left and right extreme of right panel
of Fig.~\ref{fig11}, respectively.
The spectrum of experimental
$^{20}$Ne states, each corresponding to this clusterisation, is shown
in the left panel. As already mentioned, the real nucleus will lie
somewhere between $x=1$ and $x=0$, though, the fit at $x=1$ is
acceptable.

\begin{table}\centering
\setlength{\extrarowheight}{1.5pt}
\caption
{\label{Ne20-param} 
Parameter values defining the $^{16}$O+$\alpha$ interaction. 
See Eq.~(11) in Paper I \cite{pap1}.}
\begin{supertabular}{>{\centering}p{10mm} p{10mm}<{\centering} 
p{9mm}<{\centering} p{9mm}<{\centering} p{10mm}<{\centering}
p{10mm}<{\centering} }
\hline
\hline
Hamiltonian & & & & &\\
\hline
$a$ & $\bar{a}$ & $\gamma$ & $a_{Clus}$ & $\bar{b}$ & $b$\\
-0.500 & 0.000 & 0.208 & 0.000 & 0.000 & -0.009 \\
$c$ & $a_C$ & $a^{(1)}_R$ & $t$ &&\\  
0.250 & 0.000 & 0.000 & 0.000 &&\\
\end{supertabular}
\begin{supertabular}{>{\centering}p{7mm} p{7mm}<{\centering} 
p{7mm}<{\centering} p{7mm}<{\centering} p{7mm}<{\centering}
p{7mm}<{\centering} p{7mm}<{\centering} p{7mm}<{\centering} }
\hline
\hline
Clusters & & & & & & \\
\hline
$\lambda_1$ & $\mu_1$ & $N_{0,1}$ & $\beta_1$ &
$\lambda_2$ & $\mu_2$ & $N_{0,2}$ & $\beta_2$ \\
0 & 0 & 0.00 & 0.00 & 0 & 0 & 0.00 & 0\\
\end{supertabular}
\begin{supertabular}{>{\centering}p{2mm} p{21mm}<{\centering} 
p{18mm}<{\centering} p{21mm}<{\centering} }
\hline
\hline
Quanta &&&\\
\hline
& $\hbar\omega$ & $n_0$ & $N$ \\
& 13.2 & 8 & 12\\
\hline
\hline
\end{supertabular}
\end{table}

\begin{figure}[htp]
\begin{center}
\scalebox{0.35}{\includegraphics*{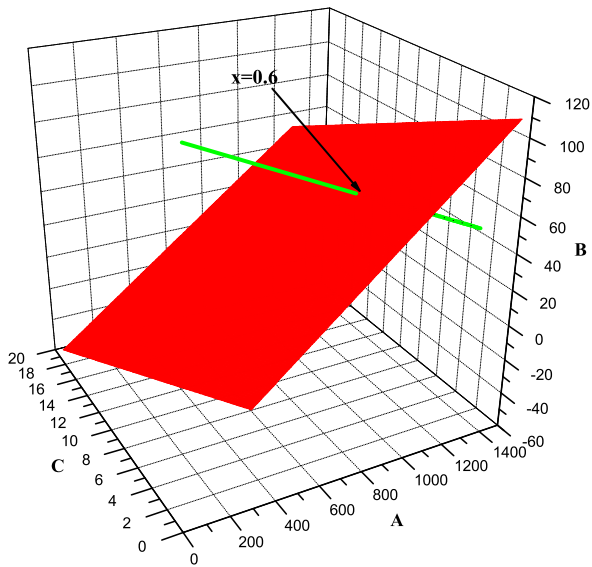}}
\end{center}
\caption{ \label{fig08}
(color online).
Part of the SACM phase space diagram, where the curve of the parameter values,
as a function of $x$, crosses the surface of phase transition. The crossing
occurs at approximately $x$ = 0.6 and happens at the surface related 
to the second-order phase transition. }
\end{figure}

In order to see if a phase transition appears, and of which order it
is, we added a curve to the $(A,B,C)$ phase space in Fig.
\ref{fig08}, depicting the transition from $x=1$ to $x=0$. The figure
shows only the relevant part of the phase space, i.e. the one where
the curve crosses the surface associated to the second-order phase
transition. This occurs at approximately the $x=0.6$ parameter
value. The actual values of $A$, $B$ and $C$ associated with this $x$
can be seen in Fig.~\ref{fig09}.  In summary, the situation in this
example corresponds to a phase transition of second order.

\begin{figure}[htp]
\begin{center}
\scalebox{0.35}{\includegraphics*{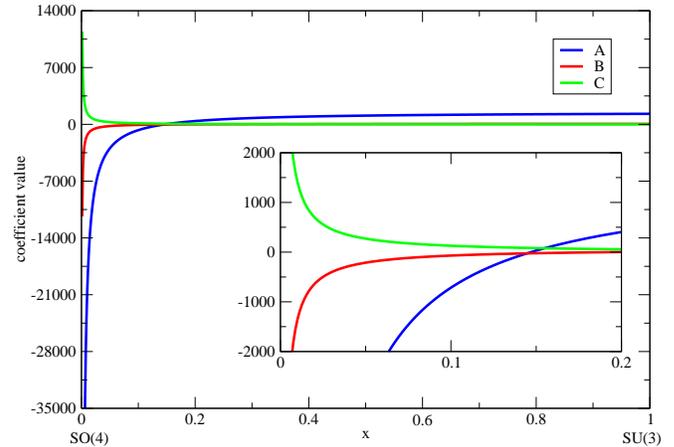}}
\end{center}
\caption{ \label{fig09}
(color online).
The parameters $A$, $B$ and $C$ as a function of $x$, considering
the $SU(3)$ to $SO(4)$ transition. The inset shows a region of large
change in $A$, $B$ and $C$ with change in $x$.
A selected range of values is shown in the inset, for a better reading.
}
\end{figure}

\begin{figure}[htp]
\begin{center}
\scalebox{0.35}{\includegraphics*{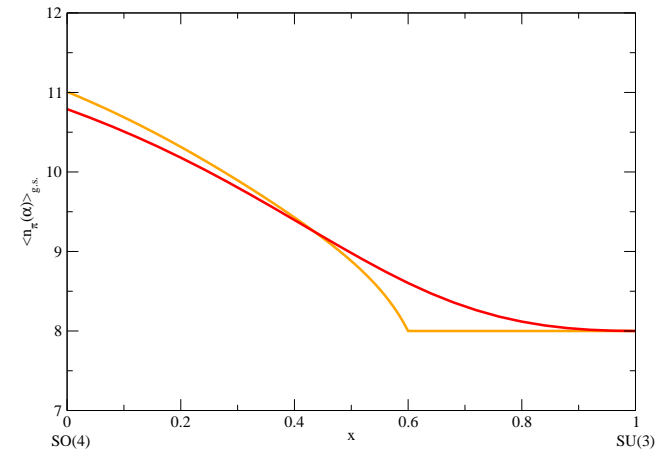}}
\end{center}
\caption{ \label{fig10}
(color online).
The expectation value of $n_\pi$ (vertical axis) as the function of
$x$.  Starting from $x=1$ the expectation value is 8, given by the
lowest possible number of $\pi$ quanta. From $x=0.6$ on we observe a
rise in the expectation value, reaching $n_\pi = 11$ at $x=0$. The
phase transition takes place at $x=0.6$.}
\end{figure}

\begin{figure}[htp]
\begin{center}
\scalebox{0.38}{\includegraphics*{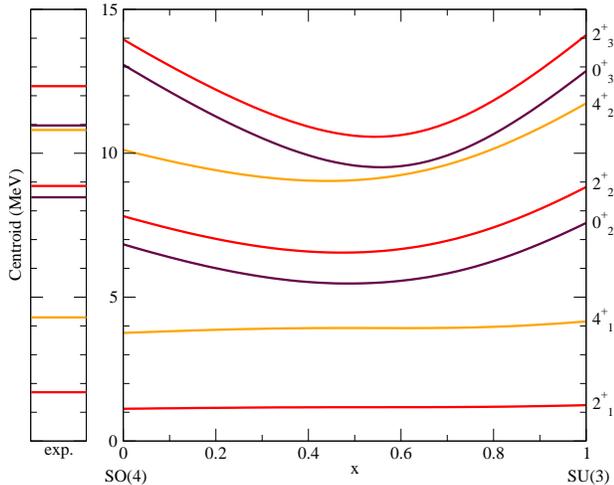}}
\end{center}
\caption{ \label{fig11}
(color online).
The lowest energy states as the function of $x$ for 
$^{16}$O+$\alpha$ $\rightarrow$ $^{20}$Ne. The spins and parities
are depicted in the legend.}
\end{figure}

Figure \ref{fig10} displays how the expectation value of
$\boldsymbol{n}_\pi$ changes as the function of $x$. The lighter
(yellow online) and darker (red online) curves depict the result of
the geometrical mapping and the numerical diagonalization,
respectively.  From $x=1$ up to the point of phase transition, the
effective $SU(3)$ limit is realized and the expectation value is equal
to the minimal number of $\pi$ bosons, i.e., $n_0=8$. Below $x=0.6$
the darker (red) curve begins to rise slightly, indicating that the structure of
the system is changing.

The right panel of Figure~\ref{fig11} shows the lowest energy states
of this clusterisation, those fitted, as the function of $x$. A smooth
behavior is observed, with an accumulation of states near
$x=0.6$. This is in accordance with the expectation value of
$\boldsymbol{n}_\pi$ in Figure~\ref{fig10}.

\subsubsection{The PACM}
\label{three-one-two}

This type of model was considered in \cite{zhang}, where a pure
schematic investigation on possible phase transitions was presented.
The model space was restricted to low $n_\pi$ excitations. No
parameter fit was applied to a physical system. We show here that the
relation to a physical system is of utmost importance and can
discriminate between physical and non-physical models.

As a first step, we tried to adjust the parameters of the model in the
$SU(3)$ limit ($x=1$ and $y=1$) and in the $SO(4)$ limit ($x=0$ and
$y=1$).  However, we already encountered severe problems in the
$SU(3)$ limit: The model space for even angular momentum starts with
$n_\pi=0$, thus the lowest states are comprised by $n_\pi=0$, $(
\lambda , \mu )=(0,0)$, which contains one $J^\pi = 0^+$ state, and by
$n_\pi = 2$, $( \lambda , \mu )=(2,0)$, which contains a $0^+$ and a
$2^+$ state.  The problems encountered are: \\
i) Since the factor in front of $\boldsymbol{L}^2$ has to be positive and
supposing that the ground state belongs to $n_\pi = 0$ and the next
excited positive parity state belongs to $n_\pi = 2$, the first
excited $0^+$ state will {\it always be lower} than the first excited
$2^+$ state. This contradicts the experimental spectrum with
$E(2_1^+)= 1.634$~MeV and $E(0_2^+)=8.7$~MeV. \\
ii) Since the $2_1^+$ state belongs to a 2$\hbar\omega$ excitation,
with $\hbar\omega~=~13.2$~MeV, the quadrupole-quadrupole interaction
has to be unnaturally strong in order to shift the energy to
1.634~MeV. This, in turn, will move very high $n_\pi$ excitations to
low energy, even below the supposed ground state with $n_\pi = 0$.\\
iii) Due to the completely different $SU(3)$ structure of the states
within the ground state ``band", one can not talk about a rotation
band.

\vskip 0.5cm
Restricting to the $SU(3)$ limit, the eigenvalues of the Hamiltonian
are given by
\begin{equation}
E = \hbar\omega n_\pi + (a-bn_\pi ) n_\pi (n_\pi +3)+\gamma L(L+1)
~~~.
\end{equation}
Here we already see that for $\gamma >0$ and a fixed $n_\pi$, higher spin
states are higher in energy. In order to adjust the $2_1^+$ and the
$0_2^+$ states to the experimental energies, the fitting routine
assigns to both the ground state and the $2_1^+$ state a {\it
  different} $n_\pi$. Using $N=20$, as a result we obtain
$n_\pi=20$ (the total number of bosons was set to be 20) for the
$0_1^+$ and the $2_1^+$ state, while the $0_2^+$ state belongs to
$n_\pi=0$.  When we change the total number of bosons, we get similar
results.  The mere fact that we have to involve states with $n_\pi =
N$ indicates that no convergence is achieved, considering that $N$
represents a cut-off value.

Similar results are also obtained when the $SO(4)$ limit is considered.
In the $SO(4)$ limit, the energy is given by
\begin{eqnarray}
E & = & \frac{c}{4}(N-\omega ) (N+\omega +2) + \gamma L(L+1)
~~~,
\end{eqnarray}
where $\omega$ refers now to the $SO(4)$ quantum number with $\omega$
= $N$, $N-2$, ..., 0 or 1. The lowest state is normally taken as
$\omega = N$, which contains $L= 0, 1, ..., \omega$.
Thus, choosing\linebreak $N=20$, the
ground-state band ($\omega = 20$) is composed of the angular momentum
states $L$ = 0, 1, ..., 20 and the first excited band with even spin
($\omega$ = 18) is given by the states $L$ = 0, 1, ..., 18. Thus, the
first excited $0^+$ state can be set at higher energies than the first
excited $2^+$ state, adjusting the parameter $c$. Everything seems to
be in order, except for the problems which the following discussion
demonstrates.

The difference with respect to the $SU(3)$ limit is that a state is a 
mixture of many basis states in $SU(3)$. We adjusted the spectrum of 
$^{20}$Ne to the $SO(4)$ limit
and confirmed that for $L=0$ the expectation value of the operator
$\boldsymbol{n}_\pi$ is given by \cite{frank}
\begin{eqnarray}
\langle \boldsymbol{n}_\pi \rangle & = & \frac{N-1}{2}
~~~.
\end{eqnarray}
The problem here is that when the {\it cut-off} is increased, the
structure of the states change: $\langle \boldsymbol{n}_\pi \rangle$
increases, implying that no convergence has been reached.  This also
implies high shell excitations, {\it if} we assume that the two
clusters are moving in a shell model mean field, which has been proven
in many microscopic calculations \cite{TE1}.
In the PACM, however, the mean field
$\hbar\omega\boldsymbol{n}_\pi$ has no specific meaning, i.e., the
parameter $\hbar\omega \rightarrow d$ can be very small. As shown
above, the spectrum can be easily adjusted within the $SO(4)$
limit. The fact that it adjusts the spectrum suggests that there must
be some truth in it. Nevertheless, the basic degrees of freedom
(clusters plus relative motion) cannot be interpreted as real clusters
or relative motion, as is done in microscopic cluster studies, but
must be in a complicated relation with them. Just what relations these
are remains a big problem.

This result demonstrates that the model shows inconsistencies and the
reasons are exposed in the arguments i) and ii) above. The
interpretational problems of the PACM have already been indicated in
\cite{huitz1}.

A possible solution to this problem is to redefine the pairing
operator as $\left[ \left( \boldsymbol{\pi}^\dagger \cdot
\boldsymbol{\pi}^\dagger \right) - R^2 \left(\sigma^\dagger
\right)^2 \right]$, i.e., the introduction of a new parameter $R^2$
which can be set proportional to $1/N$. In this way, the $N$
dependence can be eliminated and physical results can be expected.
This procedure was adopted in \cite{bijker}, where the vibron model
was extended to three clusters describing the $^{12}$C nucleus as an
oblate symmetric top.
Though, in \cite{bijker} the $SO(7)$ limit in a $U(7)$ algebraic model,
which describes three clusters, is considered, the same ideas can be
translated here.

\subsection{One spherical and one deformed cluster:
$^{20}$Ne+$\alpha$ $\rightarrow$ $^{24}$Mg}

This is the first example where the cluster part has a structure owing
to the deformed $^{20}$Ne.

\subsubsection{The SACM}
\label{three-two-one}

In order to analyze the transitions between the $SU(3)$, $SO(4)$ and
$SO(3)$ limits, first the parameters were fixed in the three
limits. The results are displayed in Table \ref{Mg24-param}.

\begin{table}\centering
\setlength{\extrarowheight}{1.5pt}
\caption
{\label{Mg24-param} 
Parameter values for the $^{20}$Ne+$\alpha$ interaction.
See Eq. (11) in Paper I \cite{pap1}.
}
\begin{supertabular}{>{\centering}p{10mm} p{10mm}<{\centering} 
p{9mm}<{\centering} p{9mm}<{\centering} p{10mm}<{\centering}
p{10mm}<{\centering} }
\hline
\hline
Hamiltonian & & & & &\\
\hline
$a$ & $\bar{a}$ & $\gamma$ & $a_{Clus}$ & $\bar{b}$ & $b$\\
-1.396 & -0.136 & 0.197 & 0.000 & -0.116 & 0.045 \\
$c$ & $a_C$ & $a^{(1)}_R$ & $t$ &&\\ 
0.470 & 0.079 & 0.053 & 0.664 &&\\
\end{supertabular}
\begin{supertabular}{>{\centering}p{7mm} p{7mm}<{\centering} 
p{7mm}<{\centering} p{7mm}<{\centering} p{7mm}<{\centering}
p{7mm}<{\centering} p{7mm}<{\centering} p{7mm}<{\centering} }
\hline
\hline
Clusters & & & & & & \\
\hline
$\lambda_1$ & $\mu_1$ & $N_{0,1}$ & $\beta_1$ &
$\lambda_2$ & $\mu_2$ & $N_{0,2}$ & $\beta_2$ \\
8 & 0 & 48.5 & 0.73 & 0 & 0 & 4.5 & 0\\
\end{supertabular}
\begin{supertabular}{>{\centering}p{2mm} p{21mm}<{\centering} 
p{18mm}<{\centering} p{21mm}<{\centering} }
\hline
\hline
Quanta &&&\\
\hline
& $\hbar\omega$ & $n_0$ & $N$ \\
& 12.6 & 8 & 12\\
\hline
\hline
\end{supertabular}
\end{table}

\begin{figure}[htp]
\begin{center}
\scalebox{0.34}{\includegraphics*{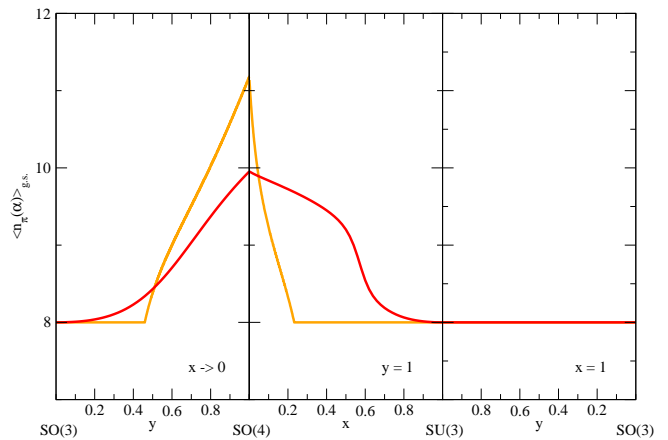}}
\end{center}
\caption{ \label{fig12}
(color online).
Expectation value of $\boldsymbol{n}_\pi$ for the three transitions
i) $SU(3)$ to $SO(4)$, ii) $SU(3)$ to $SO(3)$ and iii) $SO(4)$ to $SO(3)$. }
\end{figure}

\begin{figure}[htp]
\begin{center}
\scalebox{0.44}{\includegraphics*{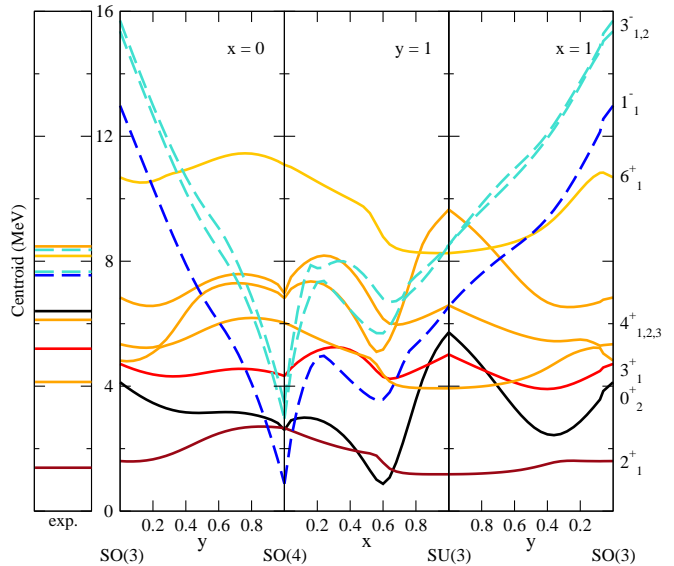}}
\end{center}
\caption{ \label{fig13}
(color online).
The lowest states in the three transition i) $SU(3)$ to $SO(4)$,
ii) $SU(3)$ to $SO(3)$ and iii) $SO(4)$ to $SO(3)$. }
\end{figure}

Figures \ref{fig12} and \ref{fig13} display the expectation value of
$\boldsymbol{n}_\pi$ and the lowest states in energy, respectively,
for all three transitions. The darker (red online) curve in
Fig.~\ref{fig12} shows the results of the numerical diagonalization,
while the lighter (yellow online) curve shows the one of the
geometrical mapping.  The results are qualitatively similar to those
encountered in the former example. The expectation value of
$\boldsymbol{n}_\pi$ starts in the $SU(3)$ limit at 8 and increases
towards the $SO(4)$ limit. The transition is smooth in the numerical
calculation, while in the geometrical mapping the transition is
pronounced well.
The transition is indicated by a sudden change in the slope at above
$x=0.2$ in the central panel of Fig.~\ref{fig12}.
The energy spectrum in Figure~\ref{fig13} does not show a particular
structure at points of phase transition.
Therein, the left-most panel
is again the experimental spectrum used, and the right three are
theoretical results as $x$ and $y$ are adjusted.
In the $SU(3)$ to
$SO(3)$ transition no phase change appears, because the global minimum
of the potential is always at $\alpha$=0. One can observe a distinct
behavior as a function of $x$ and $y$ between the states with positive
and negative parity, marked as solid and dashed curves,
respectively. The latter are more sensitive to the change in $x$ and
$y$.

Similar properties can be observed in the other limits,
i.e., see the leftmost panel in Fig.~\ref{fig12}. In the rightmost 
panel of Fig.~\ref{fig12} no phase transition occurs.
The energy spectrum also shows an accumulation of states at 
low energy at the
point of phase transition.

\subsubsection{The PACM}

Here we find similar inconsistencies with respect to the model space
as in the case of $^{20}$Ne in \ref{three-one-two}.  The cluster irrep
of $^{20}$Ne is (8,0), while the relative oscillation irreps are
$(n_\pi ,0)$, with $n_\pi$ = 0,1,2,...  Restricting to positive-parity
states only, the lowest energy model space, in the $SU(3)$ limit,
consists of $(8,0)$ at 0$\hbar\omega$, and (10,0) and (6,2) for the
2$\hbar\omega$ excitation. The ground-state band is a (8,0) irrep and
the lowest $K=2$ band is the (10,0) irrep at 2$\hbar\omega$. Even if
we change $\hbar\omega$ to an arbitrary small parameter $a_1$, in
order to bring the 2$\hbar\omega$ states down in energy, the internal
structure of the ground state band is not what we expected, namely
(8,4). Also there is no B$(E2)$ transition between the ground state
band and the $K=2$ band, because they belong to different irreps. (See
the discussion in Subsection 2.1 in Paper I \cite{pap1}.)

For the $SO(4)$ limit one obtains a satisfactory fit, but again with
the fact that the expectation value of the number operator
$\boldsymbol{n}_\pi$ depends on the cut-off $N$.

For these reasons, we do not present figures
of the spectra and expectation values.

\section{Conclusions}
\label{four}

Phase transitions were investigated in two algebraic cluster models,
one of which observed the Pauli exclusion principle between the
nucleons of the individual clusters (SACM), while the other (PACM) did
not. This analysis was based on the results of a previous work
\cite{pap1}, in which the geometric mapping of the two models had been
performed using the coherent state formalism, leading to appropriate
potential energy surfaces. The dynamical symmetries of the SACM and
PACM had also been identified in \cite{pap1}, and in the present
analysis special attention was paid to transitions between phases
associated with the $SU(3)$, $SO(3)$ and $SO(4)$ dynamical
symmetries. The potential energy surfaces depended on the parameters
appearing in the Hamiltonian shared by the SACM and PACM, including
also the $x$ and $y$ variables controlling the transitions between the
three dynamical symmetries. The phase space was reparameterized in
terms of three parameters $A$, $B$ and $C$.

In the case of the PACM, the potential energy surface was a sextic
oscillator in the intercluster distance variable, while for the SACM
the potential shape was more complex due to the restrictions enforced
by the Pauli principle. The potential energy surface typically
contained up to two minima, one spherical and one deformed.  The
analysis identified both first- and second-order phase transitions for
the PACM and the SACM, while in the latter case a critical line was
also found.

The results were illustrated with numerical studies on the
$^{16}$O+$\alpha$ and $^{20}$Ne+$\alpha$ systems, which correspond to
two spherical clusters and to one spherical and one deformed cluster,
respectively. The $SU(3)$ limit was found to be the most appropriate
one in reproducing the data of the cluster systems.  Clear phase
transitions were identified in the $x$ parameter controlling the
transition between the $SU(3)$ and $SO(4)$ limits.  It was found that
the PACM led to energy spectra that are rather different from the
observed physical ones.

\section*{Acknowledgments}

We gratefully acknowledge financial help from DGAPA, from the National
Research Council of Mexico (CONACyT), OTKA (grant No. K72357), and
from the MTA-CONACyT joint project. POH acknowledges very useful
discussions with Octavio Casta\~nos (ICN-UNAM), related to the
differences in phase transitions in finite systems to the use of
coherent states.
The authors are also thankful to J\'ozsef Cseh for illuminating 
discussions on the subject.

\end{document}